\documentclass[twocolumn,aps,showpacs,floatfix]{revtex4}
\usepackage{epsf}
\usepackage{dcolumn}
\begin{document}
\title{Bifurcations and Patterns in Compromise Processes}
\author{E.~Ben-Naim}
\email{ebn@lanl.gov}
\affiliation{Theoretical Division and Center for Nonlinear Studies, 
Los Alamos National Laboratory,  Los Alamos, New Mexico, 87545}
\author{P.~L.~Krapivsky}
\email{paulk@bu.edu}
\affiliation{Center for BioDynamics, Center for Polymer Studies, 
and Department of Physics, Boston University, Boston, MA, 02215}
\author{S.~Redner}
\email{redner@bu.edu}
\affiliation{Center for BioDynamics, Center for Polymer Studies, 
and Department of Physics, Boston University, Boston, MA, 02215}

\begin{abstract}
We study an opinion dynamics model in which agents reach compromise
via pairwise interactions.  When the opinions of two agents are
sufficiently close, they both acquire the average of their initial
opinions; otherwise, they do not interact. Generically, the system
reaches a steady state with a finite number of isolated, 
noninteracting opinion clusters (``parties'').  As the initial opinion
range increases, the number of such parties undergoes a periodic
sequence of bifurcations.  Both major and minor parties emerge, and
these are organized in alternating pattern.  This behavior is
illuminated by considering discrete opinion states.
\end{abstract}


\maketitle

\section{Introduction}
 
In a society, people typically have a wide range of opinions.
However, individual opinions on a particular issue are not static but
rather evolve due to the influences of acquaintances or other external
factors.  In principle, opinions could evolve {\it ad infinitum},
consensus could emerge, or a population could reach a state that
consists of a finite set of distinct opinion clusters, or ``parties''.

It is natural to discuss this process within the framework of
interacting particle systems \cite{weid,tml85}.  A classic example is
the voter model where agents, who possess two possible opinions, adopt
the state of a randomly selected neighbor \cite{tml85,tml99,voter1}.
Individual opinions evolve until consensus is eventually reached, and
the probability that a given opinion ultimately wins is equal to the
initial fraction of agents with that opinion \cite{voter1}.  Several
other Ising-type opinion models, incorporating more realistic
features, have been proposed recently
\cite{galam,szn,victor,holyst,stauffer,galam02}.

In this paper, we study the compromise model, a simple model for the
evolution of opinions in a heterogeneous population
\cite{david,opinion}.  To account for the diversity of the population,
the opinion is either a real-valued variable or a discrete variable
with many states \cite{F74,axelrod,CMV}.  To mimic the natural human
tendency for reaching a fair compromise, in an interaction between two
agents, both acquire the average of their initial opinions.  Last, to
incorporate self confidence or conviction in one's own opinion,
interactions between agents whose opinion difference is larger than
some threshold are forbidden.

Monte Carlo simulations of the compromise model have shown that either
consensus or diversity can arise, depending on system parameters
\cite{david,opinion}.  Here, we investigate the compromise model using
numerical integration of the governing rate equations for continuum
opinions and analytical solutions for discrete opinions.  Numerical
integration is more efficient than direct numerical simulation and
provides better resolution of the time dependent and steady state
behaviors.

We find that the compromise model exhibits a rich behavior.  In the
long-time limit, the system condenses into a finite set of
equally-spaced opinion clusters (parties), with the population in
adjacent clusters alternating between two values that differ by 4
orders of magnitude.  As the initial range of opinions grows, the
number of parties increases via a periodic sequence of bifurcations.
The corresponding period governs the basic features of the emergent
structure, namely, the size of the major clusters, and their
separation.  Near bifurcation points, the size of minor clusters
vanishes algebraically, and we provide a heuristic explanation for
this behavior.

Underlying the compromise model is a stochastic averaging process.
Closely related averaging processes naturally arise in diverse
systems, including one-dimensional inelastic collisions \cite{bk,bal},
dynamics of headways in traffic flows \cite{ff,kg}, mass transport
\cite{rm}, force fluctuations in bead packs \cite{force}, wealth
exchange processes \cite{melzak,sps}, and the Hammersley process
\cite{Hammersley,aldous}.  While our findings are discussed in the framework of
opinion dynamics, they may very well be relevant in these different
contexts.

In Sec. II, we describe the numerical integration of the rate
equations for the opinion probability density and the resulting
bifurcations.  In Sec.~III, we examine systems with a finite number
$N$ of discrete and equally-spaced opinions.  When $N$ is relatively
small, these systems can be treated analytically, thereby illuminating
the behavior in the continuum case.  Generally, consensus is reached
for small enough $N$, while a state with several distinct
non-interacting clusters is reached for large $N$.  We conclude in
Sec.~IV.

\section{the Continuum Version}

In the continuum version of the compromise model, each agent is initially
assigned an opinion $x$ from some specified distribution.  Randomly selected
pairs of agents undergo sequential interactions.  Such interactions are
restricted to agents whose opinion difference lies below a threshold that is
set to unity without loss of generality.  When agents with opinions $x_1$ and
$x_2$ interact, both acquire the average opinion:
\begin{equation}
\label{ave}
(x_1,x_2)\to \left({x_1+x_2\over 2},{x_1+x_2\over 2}\right)
\qquad |x_2-x_1|<1,
\end{equation}
while if $|x_2-x_1|>1$, no interaction occurs.  This model is essentially
identical to that of Refs.~\cite{david,opinion}.

Let us denote by $P(x,t)\,dx$ the fraction of agents that have
opinions in the range $[x,x+dx]$ at time $t$.  The distribution
$P(x,t)$ evolves according to the rate equation
\begin{eqnarray}
\label{pxt}
{\partial\over \partial t}P(x,t)&=&\int\int\limits_{|x_1-x_2|<1}
dx_1dx_2P(x_1,t)P(x_2,t)\nonumber\\
&\times&\left[\delta\left(x-{x_1+x_2\over 2}\right)-\delta(x-x_1)\right].
\end{eqnarray}
The quadratic integrand reflects the binary nature of the interaction
and the gain and loss terms reflect the process (\ref{ave}).  This
basic dynamical rule conserves the total mass and the mean opinion.
That is, $M_0$ and $M_1$, the first two moments of the opinion
distribution are conserved, where $M_k(t)\equiv\int dx\,x^k\,P(x,t)$
is the $k$th moment of the distribution. We restrict our attention to
flat initial distributions $P_0(x)\equiv P(x,0)=1$ for $x\in
[-\Delta,\Delta]$. Our goal is to determine the nature of the
final state $P_\infty(x)\equiv P(x,\infty)$.

When all agents interact, namely, when $\Delta<1/2$, the rate equations are
integrable \cite{bk,bal}. In particular, the second moment obeys $\dot
M_2+M_0M_2/2=M_1^2$, where the overdot denotes time derivative.  Using
$M_1=0$, we find that the second moment vanishes exponentially in time:
\begin{equation}
\label{M2}
M_2(t)=M_2(0)\,e^{-M_0t/2},
\end{equation}
with $M_0=2\Delta$.  Thus all agents approach the center
opinion and the system eventually reaches the consensus
\begin{equation}
\label{Pinf}
P_\infty(x)=M_0\delta(x).
\end{equation}
The distribution $P(x,t)$ approaches the localized state (\ref{Pinf})
in a self-similar fashion, $P(x,t)\simeq {2M_0\over \pi
w}\left(1+z^2\right)^{-2}$ with variance $w=M_2^{1/2}/M_0$ and scaling
variable $z=x/w$ \cite{bal}.

\begin{figure}
\centerline{\epsfxsize=7.6cm\epsfbox{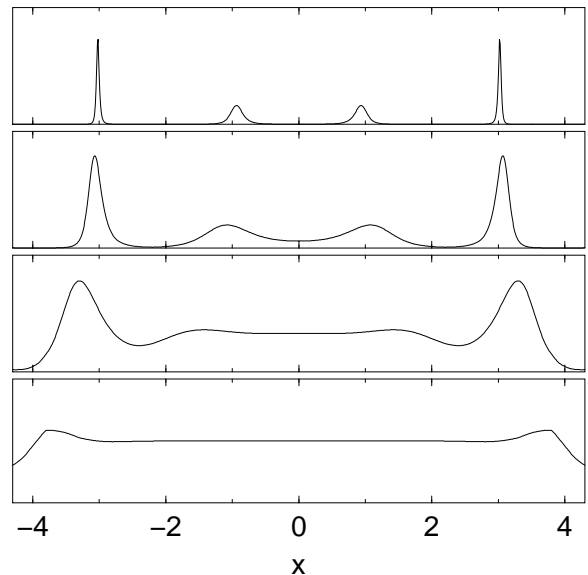}}
\caption{Evolution of the opinion distribution for $\Delta=4.3$ 
where four major clusters ultimately arise (see Fig.~2).  
Shown is $P(x,t)$ versus $x$ for times $t=0.5$ (bottom), $3$, $6$, and $9$
  (top).}
\label{evol}
\end{figure}

For larger values of $\Delta$, the opinion distribution does not
condense into a single cluster, but rather the distribution evolves
into ``patches'' that are separated by a distance larger than one.
This behavior results from an instability that propagates from the
boundary toward the center (Fig.~\ref{evol}).  Once each patch is
isolated, it then separately evolves into a delta function as in the
$\Delta<1/2$ case.  The final distribution consists of a series of
non-interacting clusters at locations $x_i$ with masses $m_i$:
\begin{equation}
\label{final}
P_\infty(x)=\sum_{i=1}^p m_i\,\delta(x-x_i)
\end{equation}
with $\sum m_i=M_0=2\Delta$ and $\sum m_ix_i=M_1=0$ to satisfy the
conservation laws.
 
Our goal is to understand basic characteristics of the final state.  How many
clusters arise?  Where are they located (in opinion space)?  What are their
masses?  As we shall see, the answers to these questions depend in a
surprisingly complex manner on the single control parameter, the initial
opinion range $\Delta$.

\subsection{Cluster Locations}

\begin{figure}
\centerline{\epsfxsize=7.6cm\epsfbox{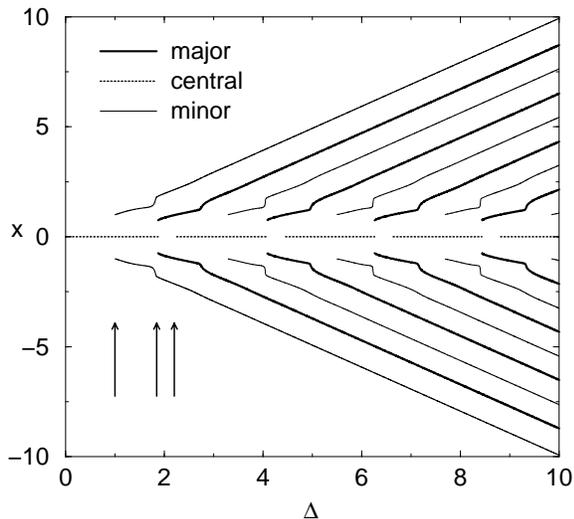}}
\caption{Location of final state clusters versus the initial opinion
  range $\Delta$.  The three types of clusters, defined in the text,
  are noted.  The vertical arrows indicate the location of the first 3
  bifurcations.}
\label{bifurcate}
\end{figure}

To determine how the final state depends on $\Delta$, we numerically
integrated the rate equation (\ref{pxt}) by discretizing $x$ into
$400\Delta$ equally-spaced states. The range $0<\Delta<10$ was
investigated using a fine mesh ($0.0025$ increments).  The rate
equations were integrated using a fourth order Adams-Bashforth method
\cite{zwillinger} up to a sufficiently long time that the probability
distribution separated into noninteracting patches.  Then, the two
conservation laws were invoked to determine the ultimate mass and
location of each patch.  The accuracy was $10^{-9}$ in $P(x,t)$.

The cluster locations exhibit a striking regularity, as seen in plotting
$x_{i}$ versus $\Delta$ (Fig.~\ref{bifurcate}).  There are three types of
clusters: major clusters (mass $M>1$), minor clusters (mass $m<10^{-2}$), and
a central cluster located exactly at $x=0$.  The number of clusters grows via
a series of bifurcations.  When $\Delta<1/2$, the final state is a single
peak located at the origin, and this situation persists as long as
$\Delta<1$.  When $\Delta$ exceeds one, two new clusters are born at the
extreme edges, $x\approx \pm\Delta$.  As $\Delta$ increases, further
bifurcations of three basic types occur:

\begin{enumerate}
\item Nucleation of a symmetric pair of clusters: $\emptyset\to
  \{-x,x\}$ with $x=1$.
\item Annihilation of a central cluster and simultaneous nucleation
  of a symmetric pair of clusters: $\{0\}\to \{-x,x\}$ with
  $x\approx 0.75$.
\item Nucleation of a central cluster: $\emptyset\to\{0\}$.
\end{enumerate}

\begin{figure}
\centerline{\epsfxsize=7.6cm\epsfbox{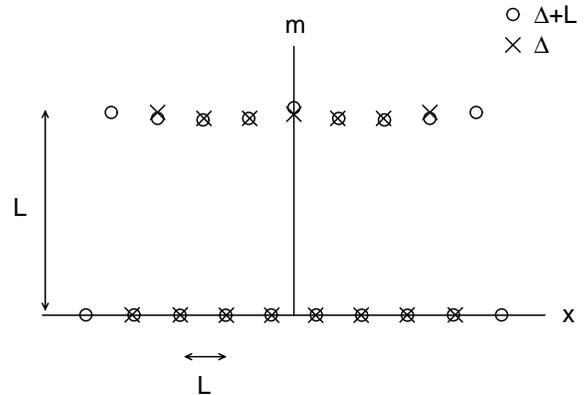}}
\caption{The masses of the final clusters versus their location for the cases
  of $\Delta\approx 7.8$ and $\Delta+L\approx 10$. The clusters in the two systems
  coincide, except that the larger system has four more clusters.  The scale
  of the inter-cluster separations and the masses of the major clusters are
  indicated.}
\label{masses}
\end{figure}

The bifurcations always occur in a periodic order:
$1,2,3,1,2,3,\ldots$.  Numerically, the first 4 generations of
bifurcations are located at
\begin{eqnarray*}
(\Delta_1,\Delta_2,\Delta_3)=\cases{(1.000, 1.871, 2.248),\cr
(3.289, 4.079, 4.455),\cr
(5.496, 6.259, 6.638),\cr
(7.676, 8.431, 8.810).}
\end{eqnarray*}
Successive bifurcation points of the same type are all separated by
the same distance: $\Delta_i(n+1)-\Delta_i(n)\to {\rm const}$.  Also,
the distance between different types of bifurcations within the same
generation eventually becomes constant. Thus the bifurcation diagram,
with all its intricate features, repeats in a periodic manner
\begin{equation} 
\label{inv}
x(\Delta)=x(\Delta+L),
\end{equation}
with period $L\approx 2.155$.  The period was estimated by
extrapolating the differences in the locations of the first few
transitions to $\infty$.

In type-1 (2) bifurcations, branches of minor (major) clusters
nucleate near the origin, and these persist for all larger $\Delta$.
As each branch evolves, notice that it exhibits a large curvature
change or a kink due to the effect of a subsequent bifurcation.  For
$|x|\agt 2$, the branch growth is practically linear with a slope
commensurate with the opinion range: $|d x/d \Delta|\to 1$.

The periodic behavior further implies that the separation between clusters
becomes constant. Moreover, when a system of size $\Delta$ is compared with a
system of size $\Delta+L$, cluster locations in the smaller system coincide
with the larger one, as shown in Fig.~\ref{masses}. The larger system,
however, contains two additional pairs of major and minor clusters. Thus, the
period $L$ governs the overall number of clusters and the separation between
them. For $\Delta\gg 1$ there are $4\Delta/L$ clusters, with neighboring
clusters separated (approximately) by distance $L/2$.

\subsection{Cluster Masses}

\begin{figure}
\centerline{\epsfxsize=7.6cm\epsfbox{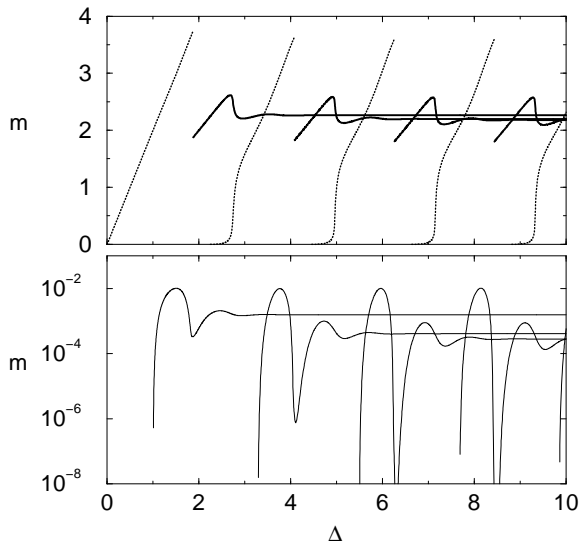}}
\caption{Cluster mass versus opinion range.   The  central clusters (periodic
  variation) and the major cluster are shown on a linear scale (top), while
  minor and central clusters are shown on a logarithmic scale (bottom).}
\label{cluster-mass}
\end{figure}

An even richer picture emerges when the cluster masses are considered.
First, the cluster masses vary periodically in the initial opinion
range, that is, $m(\Delta)=m(\Delta+L)$, as seen in Fig.~\ref{cluster-mass}.
Second, clusters are organized in an alternating major-minor pattern
(see Figs.~\ref{bifurcate} and \ref{masses}).  For large $\Delta$,
each cluster mass approaches a constant value.  The major clusters,
which contain nearly the entire mass in the system, saturate at a
value equal to the period, $M\to L$.  The masses of the minor clusters
approach a much smaller level: $m\to 3\cdot 10^{-4}$ (see
Fig.~\ref{cluster-mass}). This minute mass implies that a sufficiently
large population is needed for minor clusters to exist.

The central cluster is special.  Its mass never becomes constant but
instead varies in a periodic manner with $\Delta$
(Fig.~\ref{cluster-mass}).  A central cluster nucleates with an
infinitesimal mass at a type-3 bifurcation, grows slowly for a while,
then it undergoes an explosive growth until its mass becomes of order
unity. Finally, its mass grows linearly with $\Delta$.  At some
threshold, the central cluster splits into two major clusters via a
type-2 bifurcation (Fig.~\ref{cluster-mass}).  This birth-and-death
pattern repeats {\it ad infinitum}.

The minor clusters exhibit two subtle features.  First, the mass of
the most extreme cluster saturates to a mass $m'$ that is
approximately one order of magnitude greater than all other minor
clusters.  Second, the mass of the minor clusters varies
non-monotonically with $\Delta$ and there is a small range
of $\Delta$, where the mass of a newly-born minor cluster suddenly
drops (Fig.~\ref{cluster-mass}) before the mass saturates to a
constant value.  We are unable to resolve whether there is a
finite gap or just a singular point where the mass vanishes.

\begin{figure}
\centerline{\epsfxsize=7.6cm\epsfbox{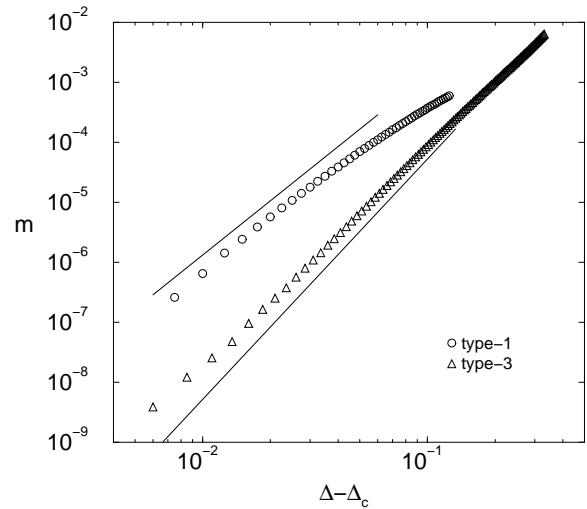}}
\caption{Critical behavior for the masses of the minor clusters at type-1 
  (top) and type-3 (bottom) bifurcations.  The straight lines have slopes 3
  and 4.}
\label{mass}
\end{figure}

At type-1 and type-3 bifurcations, new clusters form, and the mass of
these nascent clusters vanishes algebraically according to
\begin{equation}
\label{rho}  
m\sim (\Delta-\Delta_n)^{\alpha_n}
\end{equation}
as $\Delta\to\Delta_n$.  The exponent depends only on the type $n$ of
the bifurcation point; numerically we find $\alpha_1\approx 3$ and
$\alpha_3\approx 4$ (Fig.~\ref{mass}).  We now give a heuristic
explanation for this behavior.

To understand the behavior near a type-1 bifurcation, consider the very first
one at $\Delta_1=1$.  Let $\Delta=1+\epsilon$ with $\epsilon\to 0$.  It is
convenient to divide the total opinion range $(-\Delta, \Delta)$ into a
central subinterval $(-1,1)$ and two boundary subintervals: $(1,1+\epsilon)$
and $(-1-\epsilon,-1)$.  Let $m(t)$ be the mass in a boundary subinterval.
Initially, $m(0)=\epsilon$. Such mass is lost due to interaction with one
half of the central subinterval.  As a result, $\dot m=-m$, which gives
\begin{equation}
\label{rho1}  
m(t)=\epsilon\,e^{-t}.
\end{equation}
On the other hand, the mass of the central subinterval gets concentrated in a
region near the origin whose spread $w(t)$ decreases with time.  At some
moment $t_f$ the separation between masses in the central and boundary
subintervals exceeds unity.  We anticipate that the mass in the boundary
subinterval converges to its center $x=1+\epsilon/2$, and hence, this
critical separation occurs when $w(t_f)\sim \epsilon/2$.  For $t\gg t_f$, the
interaction between the two subintervals stops and the mass of the emerging
minor cluster freezes at $m_f\sim \epsilon\,e^{-t_f}$.

The spread $w(t)$ can be estimated by noting that to zeroth order in
$\epsilon$, (i) the central subinterval is not affected by the boundary
subintervals, and (ii) eventually, almost all agents are within the
interaction range.  Therefore, the asymptotic behavior is the same as in the
case $\Delta<1/2$, and the spread follows directly from the second moment
(\ref{M2}), $w(t)\sim M_2^{1/2}\sim e^{-t/2}$ since $\Delta=1+\epsilon\cong
1$. Using the stopping criteria, $w(t_f)\sim e^{-t_f/2}\sim \epsilon$, the
final minor cluster mass is $m_f\sim \epsilon\,e^{-t_f}\sim \epsilon^3$,
leading to $\alpha_1=3$.

Consider now a type-3 bifurcation that occurs at some $\Delta_3$.  We write
$\Delta=\Delta_3+\epsilon$ and adapt our previous argument.  Let $m(t)$ be
the mass of the newly-formed central cluster, and let $M$ be the final mass
of the two major clusters surrounding it.  We have $\dot m=-m M$, since the
central cluster interacts with half of the mass $M$ on either sides.
Therefore
\begin{equation}
\label{rho3}  
m(t)\sim e^{-Mt}.
\end{equation}
In contrast to Eq.~(\ref{rho1}) where the amplitude was of order $\epsilon$,
the amplitude in Eq.~(\ref{rho3}) is of order unity.  This arises because the
range of opinions that contributes to the ultimate central cluster is of the
order of the interaction range.  Now the argument proceeds as before.  The
width of the large cluster varies as $e^{-Mt/4}$.  The condition for the
central cluster and its neighbors to decouple is $e^{-Mt_f/4}\sim \epsilon$.
At this point, we have $m_f\sim e^{-Mt_f}\sim \epsilon^4$, resulting in the
exponent $\alpha_3=4$.

The heuristic argument we have presented is consistent with the
extremely small mass of the minor clusters.  For large $\Delta$, the
system is governed by the parameter $\tilde\epsilon\equiv{1\over
2}L-1\approx 0.08$, the excess between the adjacent cluster separation
and the interaction range. This parameter essentially plays the role
of $\epsilon$, the small distance from a bifurcation point.  The two
extreme minor clusters evolve according to the mechanism that led to
the $\epsilon^3$ behavior near a type-1 bifurcation. Thus, their mass
can be estimated $m'\sim \tilde\epsilon^3\approx 5\cdot 10^{-4}$. On
the other hand, minor clusters in the bulk evolve according to the
mechanism that led to the $\epsilon^4$ behavior near a type-3
bifurcation.  Accordingly, their mass is estimated by $m\sim
\tilde\epsilon^4\approx 4\cdot 10^{-5}$, again a reasonable value.

\section{The Discrete Version}

Often, one faces a choice among a finite set of options, so it is
natural to consider a discrete version of the compromise model.  While
interesting on its own, the discrete model also enables us to
illuminate many qualitative aspects of the behavior in the continuum
case.  Discrete systems are governed by a finite set of non-linear
rate equations, so explicit solutions are generally impossible.
Nevertheless, we can gain considerable insight by investigating small
systems, using stability analysis and related tools from theory of
ordinary differential equations \cite{strogatz}.

In the discrete version, each agent can take on an opinion from a set of $N$
equally-spaced values.  To impose an interaction threshold and also to ensure
that the outcome of an interaction remains within the state space, two agents
interact as follows: (a) If the opinion difference is greater than two, there
is no interaction.  (b) If the difference equals two, the agents reach a fair
compromise and each takes on the average opinion value.  (c) If the opinion
difference equals one, nothing happens.

We label the opinion states as $i=1,2,\ldots, N$, so schematically, in a
compromise event $(i-1,i+1)\to (i,i)$. Denote by $P_i(t)$ the fraction of the
population that has opinion state $i$ at time $t$.  For general $N$ the
fractions $P_i(t)$ obey the rate equations
\begin{eqnarray}
\label{rate-eq}
\dot P_i= 2P_{i-1}P_{i+1}-P_i(P_{i-2}+P_{i+2}).
\end{eqnarray}
This equation formally applies for $i$ at least 2 spacings away from the
boundaries (at $1$ and $N$).  Setting $P_{-1}=P_0=P_{N+1}=P_{N+2}\equiv 0$ 
in Eq.~(\ref{rate-eq}), yields the governing equations near the boundaries:  
$\dot P_1=-P_1P_3$, $\dot P_2=2P_1P_3-P_2P_4$, $\dot P_N=-P_{N}P_{N-2}$, and
$\dot P_{N-1}=2P_NP_{N-2}-P_{N-1}P_{N-3}$.  Again, the fractions $P_i(t)$
satisfy two conservation laws:
\begin{eqnarray}
\label{sums}
\sum_{i=1}^N P_i=1, \qquad
\sum_{i=1}^N iP_i=A,
\end{eqnarray}
with $1\leq A\leq N$.  The former (latter) reflects conservation of
the total population (opinion).  As a result, there are $N-2$
independent variables for an $N$-state system.

\subsection{Typical Behavior}

Equations (\ref{rate-eq}) are non-linear and therefore for $N\geq 4$
they cannot be solved to obtain explicit formulae for $P_i(t)$.
However, the qualitative behavior can be still understood.  For
example, Eqs.~(\ref{rate-eq}) admit only the simplest type of
attractors -- fixed points -- while limit cycles are impossible.  We
illustrate this by analyzing small values of $N$ to highlight the new
qualitative features that arise as $N$ increases.

\subsubsection{Isolated fixed points}

For $N=3$, there is a single fixed point located at 
\begin{eqnarray*}
\cases{(2-A,A-1,0) & when $1\leq A\leq 2$,\cr
       (0,3-A,A-2) & when $2\leq A\leq 3$.}
\end{eqnarray*}
This point is {\em stable}.  Asymptotically, it is approached exponentially
fast in time; {\it e.g.}, for $1<A<2$ one finds $P_3\propto e^{-(2-A)t}$.  An
exception arises for the symmetric initial condition ($A=2$) when the final
central state $(0,1,0)$ is approached algebraically in time: $P_1=P_3\to
t^{-1}$.

For $N=4$, there is also a single stable fixed point located at 
\begin{eqnarray*}
\cases{(2-A,A-1,0,0) & when $1\leq A\leq 2$,\cr
       (0,3-A,A-2,0) & when $2\leq A\leq 3$,\cr
       (0,0,4-A,A-3) & when $3\leq A\leq 4$.}
\end{eqnarray*}
Additionally, there is a fixed point $({4-A\over 3},0,0,{A-1\over 3})$ that
is always unstable.  The stable fixed point is approached exponentially in
time.

\subsubsection{Lines of fixed points}

For $N=5$, some fixed points are no longer isolated but instead they form
{\em lines}. Indeed, Eqs.~(\ref{rate-eq}) admit fixed points of the generic
forms $(P_1^*,P_2^*,0,0,P_5^*)$ and $(P_1^*,0,0,P_4^*,P_5^*)$ that are stable
when, respectively, $P_1^*>3P_5^*$ or $P_5^*>3P_1^*$.  Recalling the
conservation laws (\ref{sums}) we can write these fixed points in the form
\begin{eqnarray}
\label{stable}
\cases{
(2-A+3P_5^*,A-1-4P_5^*,0,0,P_5^*),\cr
(P_1^*,0,0,5-A-4P_1^*,A-4+3P_1^*)}
\end{eqnarray}
and obviously the fixed points form lines.  The fixed points
(\ref{stable}) from the first line are stable when $1+4P_5^*<A<2$, the
fixed points (\ref{stable}) from the second line are stable when
$4<A<5-4P_1^*$.  There is also a stable isolated fixed point located at
\begin{eqnarray}
\label{single}
\cases{(0,3-A,A-2,0,0) & when $2<A\leq 3$,\cr
       (0,0,4-A,A-3,0) & when $3\leq A<4$.}
\end{eqnarray}

For $2<A<4$, every initial condition is in the basin of attraction of the
isolated fixed point (\ref{single}).  Therefore we know the fate of the
system without explicitly solving the rate equations.  This statement tacitly
assumes that a trajectory does not approach a limit cycle or other
complicated attractor; this will be justified later.  In the complementary
range $A\in (1,2)\cup(3,4)$ the trajectories approach one of the stable fixed
points in (\ref{stable}).  For example, if $1<A<2$, the final state is
$(2-A+3P_5^*,A-1-4P_5^*,0,0,P_5^*)$ for some $P_5^*\in
\left(0,\frac{A-1}{4}\right)$.  To determine which fixed point is actually
reached depends not only on the initial average opinion
$A=\sum iP_i(0)$ but also on other details of the
initial condition and requires a complete solution.  Qualitatively, for an
initial state that is central in character ($2<A<4$), the final occupation
fractions are concentrated in a single central cluster consisting of two
adjacent sites.  Conversely, for an initial state that is biased toward one
extreme, the final state consists of two extremal clusters.

The above elementary examples demonstrate a simple principle. The rate
equations have multiple stable fixed points. Each stable fixed point is a
basin of attraction for some region in the space of initial conditions.  The
dynamics determine which stable fixed point is eventually approached. In the
continuous version, a similar situation occurs where there are enormously
many steady states of the form (\ref{final}).  Moreover, we see how depending
on the initial conditions, the system can reach a single central cluster or
two off-center clusters.

In the remainder of this subsection we consider symmetric situations,
$P_i=P_{N+1-i}$.  For an $N$-state system, we can choose $1,2,\ldots,
\lceil N/2\rceil$ independent states, where $\lceil N/2\rceil$ is the
smallest integer that is larger than or equal to $N/2$.  Conservation
of population diminishes the number of independent variables by one,
while the second conservation law is redundant as $A=(N+1)/2$ always.

\subsubsection{Explicitly solvable case}

\begin{figure}
\centerline{\epsfxsize=5cm\epsfbox{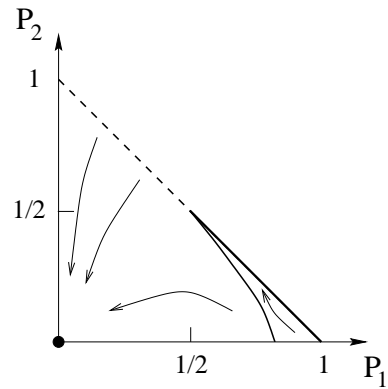}}
\caption{Schematic $(P_1,P_2)$ phase plane for the symmetric 6-opinion
  system.  Shown are the isolated fixed point (dot), the line of fixed points
  (heavy line -- dashed for unstable and solid for stable), and the
  separatrix (dotted) that demarcates the basins of attractions of these two
  sets.}
\label{triangle}
\end{figure}

For $N=6$, the rate equations (\ref{rate-eq}) are exactly solvable and the
solution neatly illustrates the features described in the previous subsection.
Using symmetry and normalization, we can treat the system in the
two-dimensional triangular domain defined by (Fig.~\ref{triangle})
\begin{equation}
\label{triangle-def}
{\cal T}=\{(P_1,P_2)| P_1\geq 0, P_2\geq 0, P_1+P_2\leq 1\}.
\end{equation}
There are two kinds of fixed points: An isolated fixed point $(0,0)$
and a line of fixed points $(P_1^*,P_2^*)$ with $P_1^*+P_2^*=1$.
Linearizing near the isolated fixed point we find that ${\bf
P}=(P_1,P_2)^T$ satisfies
\begin{eqnarray}
\label{linear6}
{d {\bf P}\over dt}= {\cal M} {\bf P}, \qquad {\rm with}\quad
{\cal M} =\pmatrix{-1&0\cr
2&-1\cr},
\end{eqnarray}
from which the origin is a degenerate stable node.  Linearizing near a fixed
point $(P_1^*,P_2^*)$ we find that it is stable iff $P_1^*>P_2^*$ and
unstable iff $P_1^*<P_2^*$.  Thus the isolated fixed point has a finite-area
basin of attraction, while every point $(P_1^*,P_2^*)$ with $P_1^*>P_2^*$ has
a basin of attraction that is a one-dimensional manifold
(Fig.~\ref{triangle}).  In principle, a two-dimensional system could have
closed orbits.  However, every closed orbit in a two-dimensional system must
enclose fixed points \cite{strogatz}.  Here, all fixed points lie on the
boundary of the phase plane $\cal T$, so closed orbits that encircle a fixed
point are impossible, thereby ruling out cycles.

The solution to Eqs.~(\ref{rate-eq}) for $N=6$ with symmetric initial
conditions is given in the Appendix.  This solution gives the following
behavior in the phase space $\cal T$ (see Fig.~5).  There is a separatrix
(\ref{separ}) that joins $(P_1,P_2)=(1/2,1/2)$ with
$\big(\sqrt{e/4},0\big)$.  The part of the phase plane to the left
of the separatrix is the basin of attraction of the isolated fixed point,
while the complementary region is the basin of attraction of the line of
fixed points $(P_1^*,P_2^*)$ with $P_1^*>{1\over 2}>P_2^*$.  These fixed
points are approached exponentially in time.  Finally, the separatrix itself
is the basin of attraction of the fixed point $(1/2,1/2)$.  In
this borderline case the relaxation is algebraic rather than exponential:
\begin{equation}
\label{separasym}
P_1-{1\over 2}\to t^{-1},\quad P_2-{1\over 2}\to -t^{-1},\quad
P_3\to 2\,t^{-2}.
\end{equation}
Both consensus and polarization are possible outcomes -- which
 actually occurs depends on the initial condition.

\subsubsection{Large $N$}

For $N\geq 7$, the systems are $\lceil N/2\rceil -1\geq 3$ dimensional, and
already the trajectories of 3-dimensional systems may exhibit a vast range of
behaviors including chaos \cite{strogatz,lorenz}.  In the present case,
however, we find that there are simply more and more fixed points, and they
appear as isolated fixed points, lines, surfaces, and higher-dimensional
sub-manifolds.

For $N=7$, the system is  three-dimensional, and the phase space is
the simplex
\begin{equation}
\label{simplex}
{\cal S}=\{(P_1,P_2,P_3)| P_j\geq 0, P_1+P_2+P_3\leq 1\}.\nonumber 
\end{equation}
For simplicity, we denote states by $(P_1,P_2,P_3,P_4)$.  The system
admits the following fixed points:
\begin{enumerate}
\item A line of fixed points $(P_1^*,P_2^*,0,0)$
  corresponding to a polarized society.
\item A line of fixed points $(P_1^*,0,0,P_4^*)$
  corresponding to a society with both centrists and extremists.
\end{enumerate}
The second case includes the central consensus state $(0,0,0,1)$ as a
special case. 

Linearizing around the fixed point $(P_1^*,P_2^*,0,0)$ we find that ${\bf
  P}=(P_3,P_4)^T$ satisfies
\begin{eqnarray}
\label{linear71}
{d {\bf P}\over dt}= {\cal M} {\bf P}, \qquad 
{\cal M} =\pmatrix{-P_1^*&2P_2^*\cr
0&-2P_2^*\cr},
\end{eqnarray}
implying that $(P_1^*,P_2^*,0,0)$ is a stable node (we tacitly assume
that $P_1^*, P_2^*>0$; when $P_1^*=2P_2^*$ this node is
degenerate). Linearizing around the fixed point $(P_1^*,0,0,P_4^*)$ we
find that ${\bf P}=(P_2,P_3)^T$ satisfies
\begin{eqnarray}
\label{linear72}
{d {\bf P}\over dt}= {\cal M} {\bf P}, \qquad 
{\cal M} =\pmatrix{-P_4^*&2P_1^*\cr
2P_4^*&-P_1^*\cr},
\end{eqnarray}
implying that $(P_1^*,0,0,P_4^*)$ is a saddle point.  Therefore the fixed
points $(P_1^*,0,0,P_4^*)$ are unstable (again it is assumed that $P_1^*,
P_4^*>0$). The two extreme fixed points $(1/2,0,0,0)$ and $(0,0,0,1)$ are
{\em neutrally} stable in the linear approximation.  Therefore one must go
beyond the linear approximation to probe the stability of consensus.
Numerically, one typically finds that the system reaches consensus ({\it
  e.g.}, starting from the uniform initial condition).  Therefore, consensus
appears to be stable.

For larger $N$, we determined the final state numerically.  To compare with
the continuum case, we start with the uniform initial condition.  Generally,
the final state consists of non-interacting clusters.  Each cluster consists
of a pair of occupied sites and clusters are separated by at least two empty
sites. We assign each cluster a mass $m$ equal to the combined occupation of
the two sites, and a position $x$ determined from a weighted average.

\begin{figure}
\centerline{\epsfxsize=7.6cm\epsfbox{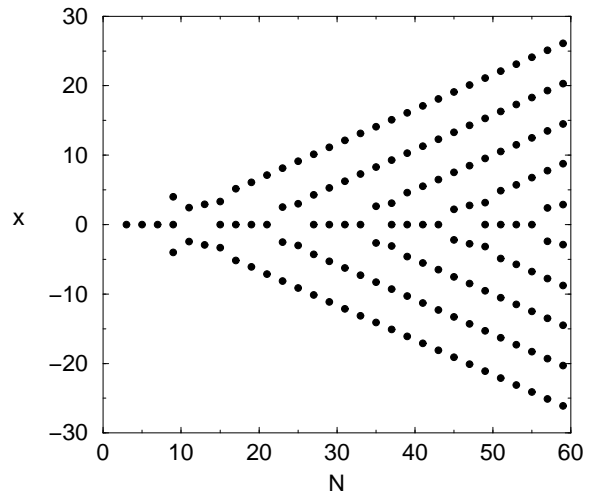}}
\caption{The location of the final clusters in the discrete model as a
  function of $N$, with $N$ odd.}
\label{m=2}
\end{figure}

As a function of $N$, the number of clusters grows via a series of
transitions, rather than bifurcations (Fig.~\ref{m=2}). The main difference
with the continuum case is that while minor clusters occasionally appear,
they do not persist in the form of minor branches. Otherwise, there are many
similarities.  Transitions involving major and central clusters are observed;
in particular, there are type-2 ($\{0\}\to \{-x,x\}$) and type-3
($\emptyset\to \{0\}$) transitions.  These transitions are arranged in a
periodic structure $2,3,2,3$, and the transition diagram is approximately
invariant under the transformation $N\to N+N_0$ with $N_0\cong 12$.  Branches
of major clusters carry almost equal masses, and remarkably, despite the
discreteness, these branches grow linearly with $N$.

\subsection{General Features}
\subsubsection{Volume Contraction}
The system of rate equations is dissipative, that is, volumes in phase
space contract under the flow.  Generally for a system of differential
equations $\dot P_j=F_j$, a volume $V(t)$ changes according to
\begin{eqnarray}
\label{volume}
{d V\over dt}= \int_S {\bf F}\cdot {\bf n}\,dS
=\int_V \nabla \cdot {\bf F}\,dV    
\end{eqnarray}
where ${\bf n}$ is the outward normal on the bounding surface $S(t)$ that
encloses $V(t)$ and $\nabla \cdot {\bf F}=\sum_j {\partial F_j\over \partial
  P_j}$.  For the infinite discrete system ${\partial F_j\over \partial
  P_j}=-P_{j-2}-P_{j+2}$ and thus the contraction rate is twice the
(conserved) total population: $-\sum_j {\partial F_j\over \partial
  P_j}=2\sum_k P_k$.  Hence volumes in phase space shrink exponentially in
time.  If we set the total population equal to 1, we get
$V(t)=V(0)\,e^{-2t}$.

For finite systems, the contraction rate is generally not a constant but
nevertheless volume still shrinks exponentially in time according to the
bounds
\begin{eqnarray}
\label{volumebound}
V(0)\,e^{-2t}\leq V(t)\leq V(0)\,e^{-t}.
\end{eqnarray}
For example, for $N=4$ the contraction rate is $\sum_k P_k=1$; therefore,
$V(t)=V(0)\,e^{-t}$.  For $N=5$, the contraction rate is $1+P_3$; for $N=6$,
the contraction rate is $1+P_3+P_4$, {\it etc}.  This is consistent with the
system evolving toward fixed points.  Nevertheless, cycles and strange
attractors are possible in volume contracting systems, with the celebrated Lorenz
system being a prime example \cite{lorenz}.

\subsubsection{Lyapunov functions}

We now demonstrate that our system has only fixed points (many of which 
are actually fixed submanifolds in phase space) by constructing a {\em
  Lyapunov function} $L\equiv L[{\bf P}(t)]$, {\it viz.}  a smooth function
that decreases along trajectories.  The existence of a Lyapunov function
rules out cycles.  Indeed, suppose that there is a periodic solution with
period $T$, then the integral $\int_0^T dt\,{d L\over dt}$ over the period
would be negative because the Lyapunov function is decreasing.  On the other
hand, the integral must be equal to zero since the trajectory returns to the
starting point.  This contradiction means that no periodic solutions can
exist.

Consider, {\it e.g.}, symmetric situations.  For $N=7$,
\begin{equation}
\label{Lyap7}
L=P_4+2P_3+4P_2+8P_1
\end{equation}
is a Lyapunov function; indeed, it satisfies
\begin{equation}
\label{Lyap7evol}
{d L\over dt}=-2P_1P_3-2P_2P_4,
\end{equation}
so the derivative is strictly negative inside the simplex ${\cal S}$
(it vanishes only on the two lines of fixed points on the boundary of
${\cal S}$). 

Generally, we can construct Lyapunov functions for all $N$.  For
instance, when $N$ is odd we write $N=2M-1$ and verify that
\begin{equation}
\label{LyapM}
L=\sum_{j=1}^M 2^{M-j}P_j,
\end{equation}
is a Lyapunov function as it satisfies 
\begin{equation}
\label{LyapMevol}
{d L\over dt}=-2P_{M-2}P_M-\sum_{j=1}^{M-3} 2^{M-2-j}P_jP_{j+2}. 
\end{equation}

\subsubsection{Negative Diffusion Instability}

In the absence of boundaries, any uniform state is a trivial solution of the
nonlinear set of the ordinary differential equations (\ref{rate-eq}). To
check the stability of the uniform state, $P_i={\rm const.}$, we treat
$i\equiv x$ as a continuum variable.  Writing $P(x,t)=1+\phi(x,t)$ with
$\phi(x,t)\ll 1$, this perturbation evolves according to
\begin{eqnarray}
\label{cont}
\phi_t+\left(\phi+{7\over 6}\phi_{xx}+{1\over 2}\phi^2\right)_{xx}=0,
\end{eqnarray}
where the subscripts denote partial differentiations.  To lowest
order, this is the diffusion equation with a negative diffusion
coefficient.  Hence, the uniform state is unstable to the perturbation
$\phi(x,t)=\exp(ikx+\lambda t)$ when $k<\sqrt{6/7}$.  Therefore,
minute details of the initial conditions are magnified, ultimately
resulting in isolated clusters.  However, the nonlinear terms in
Eq.~(\ref{cont}) eventually counter the instability.

\section{Discussion}

The interplay between compromise and conviction leads to intriguing
opinion dynamics.  The system ultimately reaches a static state that
consists of a finite number of noninteracting opinion clusters, and
the number of these clusters increases via an infinite sequence of
self-similar bifurcations as the opinion range increases.  In the bulk
of the system, clusters are organized in a periodic lattice of
alternating minor and major clusters.  A central cluster may or may
not exist, and its size exhibits a complex periodic behavior.

As a model of mathematical sociology, the compromise model is
appealing in its simplicity, yet its behavior is familiar in everyday
experience.  A political system may or may not contain a centrist
party.  Alternatively, it may consist of two (or more) off-center
parties.  Furthermore, the existence of marginal parties halfway
between two major ones is also reasonable.  Artificial features of the
model, such as the identical separation between parties, can be easily
circumvented by introducing heterogeneities.  For example, since
different agents may have different levels of conviction, it may be
natural to have interaction thresholds that are specific to each
individual.

As a dynamical system, the compromise model exhibits the simplest
types of attractors, namely, fixed points that are either sinks or
saddles.  In the discrete case, we constructed Lyapunov functions and
also established that limit cycles and strange attractors are
impossible. This conclusion extends to the continuum case. The second
moment decreases monotonically with time and hence, it is a Lyapunov
functional.  Generally, each stable fixed point is the basin of
attraction of some region in the space of initial conditions. In other
words, the rate equations map an initial state into a final
state. Given the large number of these states, it is not obvious how
to characterize such a map. One practical approach is to obtain
statistical properties of the final state by averaging over all
possible initial conditions.  In the discrete case, we find that
starting from a random initial state ($\{P_i(0)\}$ randomly chosen in
the $N$-dimensional hypercube) the distribution of cluster masses and
the distribution of the separations between clusters in the final
state are independent of the system size, for large enough systems.

There are important questions concerning robustness of the bifurcation
diagram with respect to variations in the dynamical rules or in the
initial conditions. We examined only the former.  When the opinion
difference between two agents is merely reduced by a fixed factor,
{\it i.e.}, they reach partial compromise, an almost identical
bifurcation diagram is found.  The effects of asymmetry in the initial
conditions, and the dependence of the location of bifurcation points
on the shape of the initial distribution deserve a careful study.

There are numerous possible generalizations of the compromise model.
One is to increase the dimension of the opinion space.  Do the final
opinions form a lattice? and if yes of what type?  Yet another open
question is the role of spatial dimension.  In the present work, we
implemented the mean-field limit where any agents equally well
interact with any other agent.  However, if agents are located at
lattice sites with interactions only between nearest neighbors,
spatial correlations are expected to emerge \cite{FKB,VKR}.  We
anticipate that for the discrete system in which $N\geq N_c(d)$, with
$N_c(d)$ depending on the spatial dimension $d$, the system will
freeze into a large number of non-interacting opinion domains.  The
case of continuum opinions is unexplored, but we expect both slow
dynamics and coarsening of opinion patterns as the final state is
approached.

\acknowledgments

We thanks Z.~A.~Daya and H.~A.~Rose for useful discussions.  This
research was supported by DOE (W-7405-ENG-36) and NSF(DMR9978902).

\appendix
\section{Solution to the Rate Equations for $N=6$}

The symmetric $N=6$ system simplifies after replacement of the
time variable $t$ by
\begin{equation}
\label{tau}
\tau=\int_0^t dt'\,P_3(t'),
\end{equation}
as the rate equations reduce to the linear system
\begin{eqnarray*}
\label{rate6new}
P_1'=-P_1,\quad
P_2'=2P_1-P_2,\quad
P_3'&=&P_2-P_1,
\end{eqnarray*}
where $'\equiv d/d\tau$.  Solving these equations we obtain
\begin{eqnarray}
\label{rate6sol}
P_1(\tau)&=&P_1(0)\,e^{-\tau},\nonumber\\
P_2(\tau)&=&\left[2P_1(0)\,\tau +P_2(0)\right]\,e^{-\tau},\\
P_3(\tau)&=&1
-\left[2P_1(0)\,\tau +P_1(0)+P_2(0)\right]\,e^{-\tau}.\nonumber
\end{eqnarray}

Depending on the initial conditions, $P_3(\tau)$ either remains
positive or it reaches zero. In the former case, $P_3(\tau)\to 1$ as
$\tau\to\infty$ and asymptotically the isolated fixed point is
reached.  To determine the approach to the fixed point in terms of
original time variable we write
\begin{equation}
\label{time}
t=\int_0^\tau {d\tau'\over P_3(\tau')},
\end{equation}
with $P_3$ given by (\ref{rate6sol}). 
Asymptotically we find
\begin{eqnarray*}
\tau&=&t-c+{\cal O}(t\,e^{-t}),\\
c&=&\int_0^\infty d\tau\,
{\left[2P_1(0)\,\tau +P_1(0)+P_2(0)\right]\,e^{-\tau}\over 
1-\left[2P_1(0)\,\tau +P_1(0)+P_2(0)\right]\,e^{-\tau}}.
\end{eqnarray*}
Substituting this into (\ref{rate6sol}) we arrive at
\begin{eqnarray*}
P_1(t)&=&\Pi_1\,e^{-t}+{\cal O}(t\,e^{-2t}),\\
P_2(t)&=&\left[2\Pi_1\,t +\Pi_2\right]\,e^{-t}+{\cal O}(t^2\,e^{-2t}),
\end{eqnarray*}
with $\Pi_1=P_1(0)\,e^{c}$, and $\Pi_2=\left[P_2(0)-2cP_1(0)\right]\,e^{c}$.

In the complementary situation, $P_3(\tau^*)=0$ at some $\tau^*$ and then it 
becomes negative.  In the limit $\tau\to\tau^*$ the physical time diverges;
see Eq.~(\ref{time}).  Therefore the range $\tau\geq \tau^*$ is physically
forbidden so that the system reaches a fixed point $(P_1^*,P_2^*)$, with
\begin{eqnarray*}
P_1^*&=&P_1(0)\,e^{-\tau^*},\\
P_2^*&=&\left[2P_1(0)\,\tau^* +P_2(0)\right]\,e^{-\tau^*}.
\end{eqnarray*}
The approach to this fixed point is exponential in time. 

The borderline between these two regimes occurs when $P_3(\tau)>0$ for all
$\tau\ne \tau^*$, i.e., the curve $P_3(\tau)$ touches the $\tau$ axis
horizontally. Thus we require {\em both}
\begin{equation}
\label{tau1}
1=\left[2P_1(0)\,\tau^* +P_1(0)+P_2(0)\right]\,e^{-\tau^*}
\end{equation}
and
\begin{equation}
\label{tau2}
2P_1(0)=2P_1(0)\,\tau^* +P_1(0)+P_2(0).
\end{equation}
The second relation gives $\tau^*=[P_1(0)-P_2(0)]/[2P_1(0)]$.
Substituting this into (\ref{tau1}) yields the separatrix
\begin{equation}
\label{separ}
{P_1(0)-P_2(0)\over 2P_1(0)}=\ln 2P_1(0).
\end{equation}

\end{document}